\documentclass[prd,showpacs,twocolumn,
amsmath,amssymb,superscriptaddress,floatfix,nofootinbib]{revtex4-1}
\usepackage[utf8]{inputenc}
\usepackage[sort&compress]{natbib}
\usepackage[normalem]{ulem}
\usepackage{lipsum} 
\usepackage{bm}
\usepackage{times}
\usepackage{amssymb,amsbsy,amsmath,amsfonts}
\usepackage{graphicx}
\usepackage{float}
\usepackage{color}
\usepackage{morefloats}
\usepackage{rotating}
\usepackage{srcltx}
\usepackage{slashed}
\usepackage{subfigure}
\usepackage{multirow}
\usepackage{verbatim}
\usepackage{hyperref}
\usepackage{tabularx}
\usepackage{adjustbox}
\usepackage{overpic}
\usepackage[dvipsnames]{xcolor}
\usepackage{nicefrac}

\usepackage{hyperref}
\hypersetup{
	colorlinks	=true,
	urlcolor	=blue,
	linkcolor	=blue,
	citecolor	=blue,
	pdftitle	={},
	pdfauthor	={Zhi-Wei Liu, Jun-Xu Lu, Ming-Zhu Liu, Li-Sheng Geng},
	pdfsubject	={$\Sigma_c\bar{D}^{(*)}$ correlation function}
	}

\begin{document}

\title{ Investigations on   the weak  decays of $D\bar{B}$ molecules       }

\author{Ming-Zhu Liu}
\affiliation{
Frontiers Science Center for Rare Isotopes, Lanzhou University,
Lanzhou 730000, China}
\affiliation{ School of Nuclear Science and Technology, Lanzhou University, Lanzhou 730000, China}

\author{Li-Sheng Geng}
\email[Corresponding author: ]{lisheng.geng@buaa.edu.cn}
\affiliation{School of Physics, Beihang University, Beijing 102206, China}
\affiliation{Peng Huanwu Collaborative Center for Research and Education, Beihang University, Beijing 100191, China}
\affiliation{Beijing Key Laboratory of Advanced Nuclear Materials and Physics, Beihang University, Beijing 102206, China }
\affiliation{Southern Center for Nuclear-Science Theory (SCNT), Institute of Modern Physics, Chinese Academy of Sciences, Huizhou 516000, China}

\date{\today}
\begin{abstract}

The decays  of  exotic states discovered experimentally   always  proceed via  the strong and electromagnetic interactions.  Recently,  a  tetraquark state with the quark content $bc\bar{q}\bar{q}$ was    predicted by Lattice QCD simulations. It is below the mass threshold of $D\bar{B}$, which can only decay via the weak interaction. In this work, based on the decay mechanism of $T_{cc}$ as a $DD^*$ molecule,  we propose that  the decays of the  $bc\bar{q}\bar{q}$ tertaquark state as a  $D\bar{B}$ molecule  proceed  via the Cabibbo-favored  weak decays of the $\bar{B}$ or $D$ meson, accompanied by the tree-level decay modes and the triangle decay modes. Our results indicate that the branching fraction of  the $D\bar{B}$ molecule  decaying into  $\pi^+ K^{-} \bar{B}^0$ is sizable, which is a good channel to observe the $D\bar{B}$ molecule in future experiments.

\end{abstract}


\maketitle

\section{Introduction}

Many new hadron states beyond  mesons made of a pair of quark and anti-quark and baryons made of three quarks in the conventional quark model, often named as exotic states,  have been discovered in recent years. Their  quark configurations  in terms of Quantum ChromoDynamics (QCD) can be either  compact multiquark state, hybrid, hadron-charmonium, or hadronic molecule~\cite{Chen:2016qju,Hosaka:2016ypm,Lebed:2016hpi,Oset:2016lyh,Esposito:2016noz,Dong:2017gaw,Guo:2017jvc,Olsen:2017bmm,Ali:2017jda,Karliner:2017qhf,Guo:2019twa,Brambilla:2019esw,Liu:2019zoy,Meng:2022ozq,Liu:2024uxn}.      Among them,  the hadronic molecular picture, where these states are composed by a pair of conventional hadrons, have been   intensively discussed, motivating  us to study the relevant hadron-hadron interactions~\cite{Junnarkar:2019equ,Cheung:2020mql,Prelovsek:2020eiw,Padmanath:2022cvl,Wilson:2023hzu,Lyu:2023xro,Chen:2022vpo} as well as explore the corresponding  few-body hadronic molecules~\cite{MartinezTorres:2009xb,Zhang:2019ykd,Wu:2022ftm,Tan:2024omp}.  There exist some  candidates of hadronic molecules,   such as  $D_{s0}^*(2317)$ and $D_{s1}(2460)$ as the $DK$ and $D^*K$ molecules,  $X(3872)$  as the $\bar{D}^*D$ molecule,  $P_{c}(4312)$,  $P_{c}(4440)$ and $P_{c}(4457)$ as the $\bar{D}^{(*)}\Sigma_c$ molecules, $T_{cc}$ as the $D^*D$ molecule, and so on~\cite{Liu:2024uxn}.  Up to now,  the molecular interpretations for the exotic states have not been firmly established,  but at least one can conclude that these states contain sizable molecular components in their wave functions.

The decay of a hadronic molecule is responsible for its  width.  According to the number of final states in the decay, the decay modes   of  hadronic molecules can contain  two-body, three-body,  or four-body, among  which the former two are rather common in hadroic molecules.  For the two-body decay mode of a hadronic molecule, the inelastic hadron-hadron potential is crucial to calculate its partial decay width.  A typical example is  that the $P_{c}(4312)$,  $P_{c}(4440)$, and $P_{c}(4457)$,  as  $\bar{D}^{(*)}\Sigma_c$ molecules,    decaying  into   $J/\psi p$ and $\bar{D}^{(*)}\Lambda_c$ can be described by  the one-boson exchange model~\cite{Xiao:2019mvs,Lin:2019qiv,Yamaguchi:2019seo,He:2019rva,Yalikun:2021bfm,Shen:2024nck} or effective field theories~\cite{Sakai:2019qph,Du:2019pij,Xiao:2019aya,Pan:2023hrk}. Due to   the large uncertainties  in the $\bar{D}^{(*)}\Sigma_c \to J/\psi p$ and $\bar{D}^{(*)}\Sigma_c \to \bar{D}^{(*)}\Lambda_c$  potentials, there are large uncertainties in their  partial decay widths as well.     As for the three-body decay mode, the decay of  a hadronic molecule proceeds via the decay of either its constituent.  A classical example is that the doubly charmed tetraquark $T_{cc}$ as a $D^*D$ bound states decaying  into $DD\pi$ and $DD\gamma$ proceeds via  the off-shell $D^*$ meson decaying into $D\pi$ and $D\gamma$~\cite{Meng:2021jnw,Ling:2021bir,Du:2021zzh,Feijoo:2021ppq,Albaladejo:2021vln,Yan:2021wdl,Fleming:2021wmk,Chen:2021vhg,Dai:2021wxi,Wang:2023ovj,Zhang:2024dth,Sun:2024wxz}.   Similarly,     it's natural to expect that the $\bar{D}^{(*)}\Sigma_c$ molecules can decay into $\bar{D}^{(*)} \Lambda_c \pi$ via the off-shell $\Sigma_c$ baryon decaying into $\Lambda_c \pi$~\cite{Xie:2022hhv}, while no significant signal is  observed in a recent analysis of LHCb Collaboration~\cite{LHCb:2024pnt}.

Due to the fact that the order of magnitude of weak decays is much smaller than those of strong and radiative decays, the weak decay of a hadronic molecule is always neglected. Moreover,  since hadronic molecules are all observed via their strong  or radiative decays, the weak decay of hadronic molecule is scarcely discussed in the literature. In Ref.~\cite{Branz:2008cb},   assuming  the $D_{s0}^*(2317)$ as the $DK$ molecule,  Branz et al., investigated the  weak decays of $D_{s0}^*(2317) \to f_0(980) X$( $X=\pi, K $,  and $\rho$ mesons)  via the triangle diagram mechanism. In Ref.~\cite{Gal:2024nbr}, the weak decay of the doubly strange dibaryon  $\Lambda \Lambda$  into a pair of nucleons $nn$ is studied via the weak decay of $\Lambda \to n \pi$. In this work, we focus on the weak decays of hadronic molecules, especially those hadronic  molecules that can only  decay  weakly.

The hadronic molecules containing the quark content $\bar{Q}\bar{Q}qq$ are  particularly good for studying  weak decays of hadronic molecules, which are intensively studied since the doubly charmed tetraquark state $T_{cc}$ is discovered.                  
Very recently, Lattice QCD simulations studied the $D\bar{B}$ interaction and found a bound state below the $D\bar{B}$ mass threshold~\cite{Alexandrou:2023cqg,Radhakrishnan:2024ihu}, denoted as $T_{cb}$,  which can only be discovered via  the weak decay modes~\cite{Karliner:2017qjm}.    In this work, we take the contact-range effective field theory(EFT) to calculate the mass of  the $D\bar{B}$  molecule, then adopt the tree-level and triangle diagram  decays of  the $D\bar{B}$  molecule to calculate its partial decay widths, which are helpful   to experimentally search for the predicted $D\bar{B}$ molecule.

\begin{figure*}[htbh]
  \centering
  \includegraphics[width=0.65\textwidth]{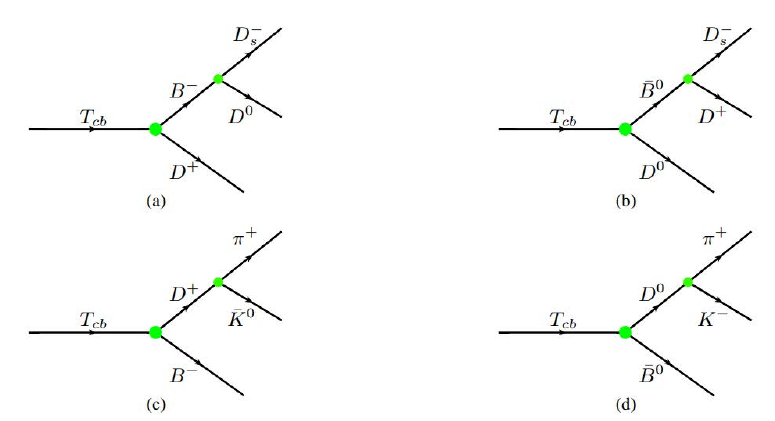}
  \caption{Tree diagrams for the decays of  $T_{cb} \to D^+ D^0 D_s^-$(a),(b),   $T_{cb} \to B^- \pi^+ \bar{K}^0$(c), and  $T_{cb} \to \bar{B}^0 \pi^+ K^-$(d).}\label{Fig:tree}
\end{figure*}

\begin{figure*}[htbh]
  \centering
  \includegraphics[width=0.62\textwidth]{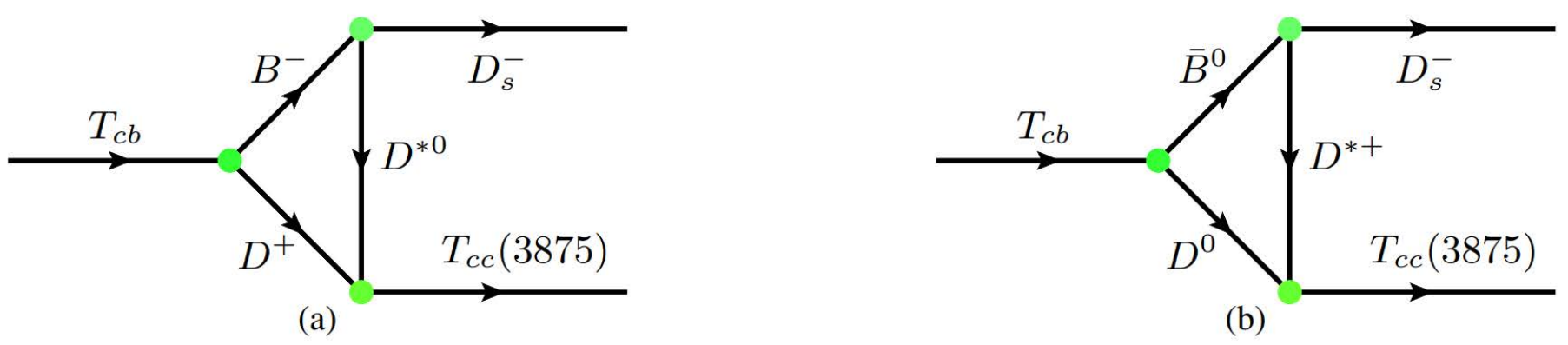}
  \caption{Triangle diagrams for the decays of $T_{cb} \to T_{cc}(3875)D_s^- $~(a),(b).}\label{Fig:tri}
\end{figure*}

This work is organized as follows. We 
briefly introduce the weak decay mechanisms of the $D\bar{B}$ molecule and the effective Lagrangian approach in Sec.~II. Numerical results and discussions are given in Sec.~III, followed by a summary in the last section.

\section{Theoretical formalism}


 Following the decay mechanism of $T_{cc}$ as a $D^*D$ molecule~\cite{Meng:2021jnw,Ling:2021bir,Du:2021zzh,Feijoo:2021ppq,Albaladejo:2021vln,Yan:2021wdl,Fleming:2021wmk,Chen:2021vhg,Dai:2021wxi,Wang:2023ovj}, we propose that  the hadronic molecule $D\bar{B}$ decay via the weak decays of the $D$ or the $\bar{B}$ meson. Considering the efficiency in experimental measurements and  focusing on the dominant branching fraction in  the $D\bar{B}$ molecule decay,   we select the   Cabibbo favored  weak decays $\bar{B} \to D^{(*)} \bar{D}_s$  and $D \to \pi K$  as the secondary decay. In  Fig.~\ref{Fig:tree},  we  illustrate the weak decays of the $D\bar{B}$ molecule via the tree diagram.  Moreover,  we describe the     $D\bar{B}$ molecule  decaying into  the $T_{cc}$ as a $DD^*$ molecule   via the triangle diagram mechanism as shown  Fig.~\ref{Fig:tri}, where $\bar{B}$ first weakly decays into $D^* \bar{D}_s$,  and then the $D^*D$ interaction dynamically generates the $T_{cc}$.

\subsection{ Effective Lagrangian }

 In this work, we employ the effective Lagrangian approach to calculate the weak decay widths. At first, we present the relevant Lagranian to be used in this work.        
The hadronic  molecule couplings to  the corresponding constituents are described by the following Lagrangians~\cite{Ling:2021bir,Liu:2022dmm} 
\begin{eqnarray}
\mathcal{L}_{T_{cb}D\bar{B}}&=&  g_{T_{cb}D\bar{B}} T_{cb}D\bar{B}, 
\\ \nonumber
\mathcal{L}_{T_{cc}D{D}^*}&=&  g_{T_{cc}D{D}^*} T_{cc}^{\mu}D {D}^*_{\mu}, 
\end{eqnarray}
where  $g$ with the specific  subscript denotes the molecule's couplings to their constituents, which are estimated in the contact-range EFT approach.   

As for the weak decays, 
the amplitudes of { $\bar{B}(k_0) \to \bar{D}_{s}(q_1) {D}^{(\ast)}(q_2)$ }  and $D(k_0) \to \pi(q_1)  K(q_2)$  have the following form~\cite{Wu:2023rrp}
 
\begin{align}\label{am3}
&\mathcal{A}(\bar{B}\to \bar{D}_{s}{D}^{\ast})= \frac{G_{F}}{\sqrt{2}}V_{cb}V_{cs} a_{1} f_{\bar{D}_{s}}\{-q_{1}\cdot \varepsilon(q_{2})   \\ \nonumber 
&(m_{ {D}^{\ast}}+m_{\bar{B}})A_{1}\left(q_{1}^{2}\right) +(k_{0}+q_{2}) \cdot \varepsilon(q_{2}) q_{1}\cdot (k_{0}+q_{2}) \\ \nonumber       & 
\frac{A_{2}\left(q_{1}^{2}\right)}{m_{ {D}^{\ast}}+m_{\bar{B}}} +(k_{0}+q_{2}) \cdot \varepsilon(q_{2}) [(m_{{D}^{\ast}}+m_{\bar{B}})A_{1}(q_{1}^2) \\ \nonumber & -(m_{\bar{B}}-m_{ {D}^{\ast}})A_{2}(q_1^2) -2m_{{D}^{\ast}} A_{0}(q_{1}^2)]  \} , \\ \nonumber
&\mathcal{A}(\bar{B} \to \bar{D}_{s} {D})=\frac{G_{F}}{\sqrt{2}}V_{cb}V_{cs} a_{1}f_{\bar{D}_{s}}(m_{\bar{B}}^2-m_{D}^2)F_{0}(q_{1}^2), \\ \nonumber
&\mathcal{A}(D \to K \pi )=\frac{G_{F}}{\sqrt{2}}V_{sc}V_{ud} a_{1}f_{\pi}(m_{D}^2-m_K^2)F_{0}(q_{1}^2), 
\end{align}
 where $a_1$ is the Wilson Coefficient,  $f_{\bar{D}_{s}}$ is  the decay constant  for the $\bar{D}_s$ meson, and $\epsilon_\mu$ denotes the polarization vector of a vector particle.  In this work, we take  $G_F = 1.166 \times 10^{-5}~{\rm GeV}^{-2}$, $V_{cb}=0.0395$, $V_{cs}=0.991$,  and $f_{\bar{D}_{s}} = 250$ MeV~\cite{FlavourLatticeAveragingGroupFLAG:2021npn,Li:2017mlw}. The form factors of  $F_{0}(t)$,  $A_{0}(t)$, $A_{1}(t)$, and $A_{2}(t)$ with $t \equiv q^{2}$ can be parameterized in the  form of
$F(t)={F(0)}/[1-a\left(t / m_{B}^{2}\right)+b\left(t^{2} / m_{ B}^{4}\right)]$.
 The values of $F_0$, $a$, and $b$ in  the transition form factors of $\bar{B}\to  {D}^{(\ast)}$   are taken from Ref.~\cite{Verma:2011yw} and shown in Table~\ref{BtoKformfactor1}.

  \begin{table}[ttt]
 \centering
 \caption{Values of  $F(0)$, $a$, $b$ in the $\bar{B} \rightarrow D^{(*)}$    transition  form factors~\cite{Verma:2011yw}. \label{BtoKformfactor1} }
 \begin{tabular}{c|ccccccccc}
 \hline\hline
  & $F_0$~~~ & $F_1$~~~ & $V$~~~ & $A_0$~~~ &  $A_1$~~~ & $A_2$   \\
 \hline  
 $F(0)^{B \to D^{(*)}}$~~~ &  0.67~~~  & 0.67~~~ & 0.77~~~ & 0.68~~~ & 0.65~~~ & 0.61~~~  \\
 $a^{B \to D^{(*)}}$~~~  & 0.63~~~ &  1.22~~~  & 1.25~~~ & 1.21~~~ & 0.60~~~ & 1.12~~~  \\
 $b^{B \to D^{(*)}}$~~~ & -0.01~~~  & 0.36~~~  & 0.38~~~ &0.36~~~ & 0.00~~~ & 0.31~~~    \\ 
\hline  \hline
 \end{tabular}
 \end{table}
 
For the weak decays of $\bar{B} \to \bar{D}_{s}{D}$ and  $D \to K \pi $,  the particles in the tree decay modes are on-shell, resulting in  the amplitudes of $\mathcal{B}(\bar{B} \to \bar{D}_{s}{D})$ and  $\mathcal{B}(D \to K \pi )$ to be the constants.    Therefore,   we further parameterise their Lagrangians as 
\begin{eqnarray}
\mathcal{L}_{\bar{B}  \bar{D}_{s}{D}}&=&  g_{\bar{B}  \bar{D}_{s}{D}} \bar{B}  \bar{D}_{s}{D}, 
\\ \nonumber
\mathcal{L}_{DK \pi }&=&  g_{DK \pi } DK \pi, 
\end{eqnarray}
where the couplings are determined by reproducing the experimental data.  With the experimental branching fractions    $\mathcal{B}(D^0 \to K^- \pi^+)=(3.947 \pm 0.030) \%$ and $\mathcal{B}(D^+ \to \bar{K}^0 \pi^+)=(3.067 \pm 0.053) \%$~\cite{Cheng:2024hdo},  we obtain the values for the couplings $g_{D^0 K^- \pi^+}=2.535 \times 10^{-6}$~GeV  and $g_{D^+ \bar{K}^0 \pi^+}=1.411 \times 10^{-6}$~GeV. Similarly, with the branching fractions of $\mathcal{B}(B^- \to D_s^- D^0)=(9.0  \pm 0. 9) \times 10^{-3}$ and $\mathcal{B}( \bar{B}^0  \to D_s^- D^+)=(7.2  \pm 0. 8) \times 10^{-3}$,  we derive the couplings of $g_{B^- D_s^- D^0}=1.182 \times 10^{-6}$~GeV  and $g_{\bar{B}^0  D_s^- D^+ }=1.098\times 10^{-6}$~GeV. To further reduce the uncertainty of the weak decay vertex $\bar{B} \to \bar{D}_s {D}^*$, we take their experimental branching fractions of $\mathcal{B}(B^- \to D_s^- D^{*0})=(8.2  \pm 1.7) \times 10^{-3}$ 
and  $\mathcal{B}(\bar{B}^0  \to D_s^- D^{*+})=(8.0  \pm 1.1) \times 10^{-3}$ to fix the effective Wilson coefficient $a_1$ as $0.93$ and $0.96$, respectively~\cite{Wu:2023rrp},  a bit smaller than that of $a_1=1.07$ at the energy  scale of $m_c$~\cite{Li:2012cfa}, which indicates that the factorisation  contribution  plays the dominant role.

 \subsection{ Contact-Range effective field theory  }

In the following, we explain how to determine the molecule couplings to their constituents in the contact-range EFT approach. These couplings are estimated by solving the Lippmann-Schwinger
equation,  
\begin{eqnarray}
 T(\sqrt{s})=[1-VG(\sqrt{s})]^{-1}V,   
\end{eqnarray}
where $G(\sqrt{s})$ is the loop function, and $V$ is the hadron-hadron  potential derived in the contact EFT approach. 
In the heavy quark limit, the contact  potentials between a pair of heavy mesons are parameterised  as~\cite{Liu:2019stu,Peng:2023lfw} 
\begin{eqnarray}
    \label{potential}
 V(I=0, \, DD^*)&=& C_a +C_b   \\ \nonumber
  V(I=0, \, D^*D^*)&=& C_a +C_b   \\  \nonumber
    V(I=0, \, D \bar{B})&=& C_a    \\  \nonumber
\end{eqnarray}
where $C_a$ and $C_b$  represent the spin-spin independent term and dependent term,  which are determined by fitting to the mass of a hadronic molecule candidate.  

To avoid the divergence of  the loop function $G(\sqrt{s})$, one  introduces a regulator of Gaussian form $e^{-2q^{2}/\Lambda^2}$ in the integral as
\begin{eqnarray}
G(\sqrt{s})=\int \frac{d^{3}q}{(2\pi)^{3}} \frac{e^{-2q^{2}/\Lambda^2}}{{\sqrt{s}}-m_{1}-m_{2}-q^{2}/(2\mu_{12})+i \varepsilon}
\label{loopfunction},
\end{eqnarray}
where $\sqrt{s}$ is  the total energy in the c.m.frame of $m_{1}$ and $m_{2}$, $\mu_{12}={m_{1}m_{2}}/(m_{1}+m_{2})$ is the reduced mass, and $\Lambda$ is the momentum cutoff. Following our previous works~\cite{Liu:2020tqy}, we take $\Lambda=0.7$~GeV in the present work.

With the  potentials given in Eq.~(\ref{potential}),  we  search for poles in the vicinity of  the $D\bar{B} $ and $D^{(*)}D^*$   mass thresholds and  then determine the 
 couplings  from the residues of the corresponding poles, 
\begin{eqnarray}
g_{i}g_{j}=\lim_{{\sqrt{s}}\to {\sqrt{s_0}}}\left({\sqrt{s}}-{\sqrt{s_0}}\right)T_{ij}(\sqrt{s}),
\end{eqnarray}
where $g_{i}$ denotes the coupling of channel $i$ to the  dynamically generated state and ${\sqrt{s_0}}$ is the pole position. 


 \begin{figure*}[htph]
\begin{center}
\begin{tabular}{cc}
\subfigure[]
{
\begin{minipage}[t]{0.46\linewidth}
\begin{center}
\begin{overpic}[scale=.36]{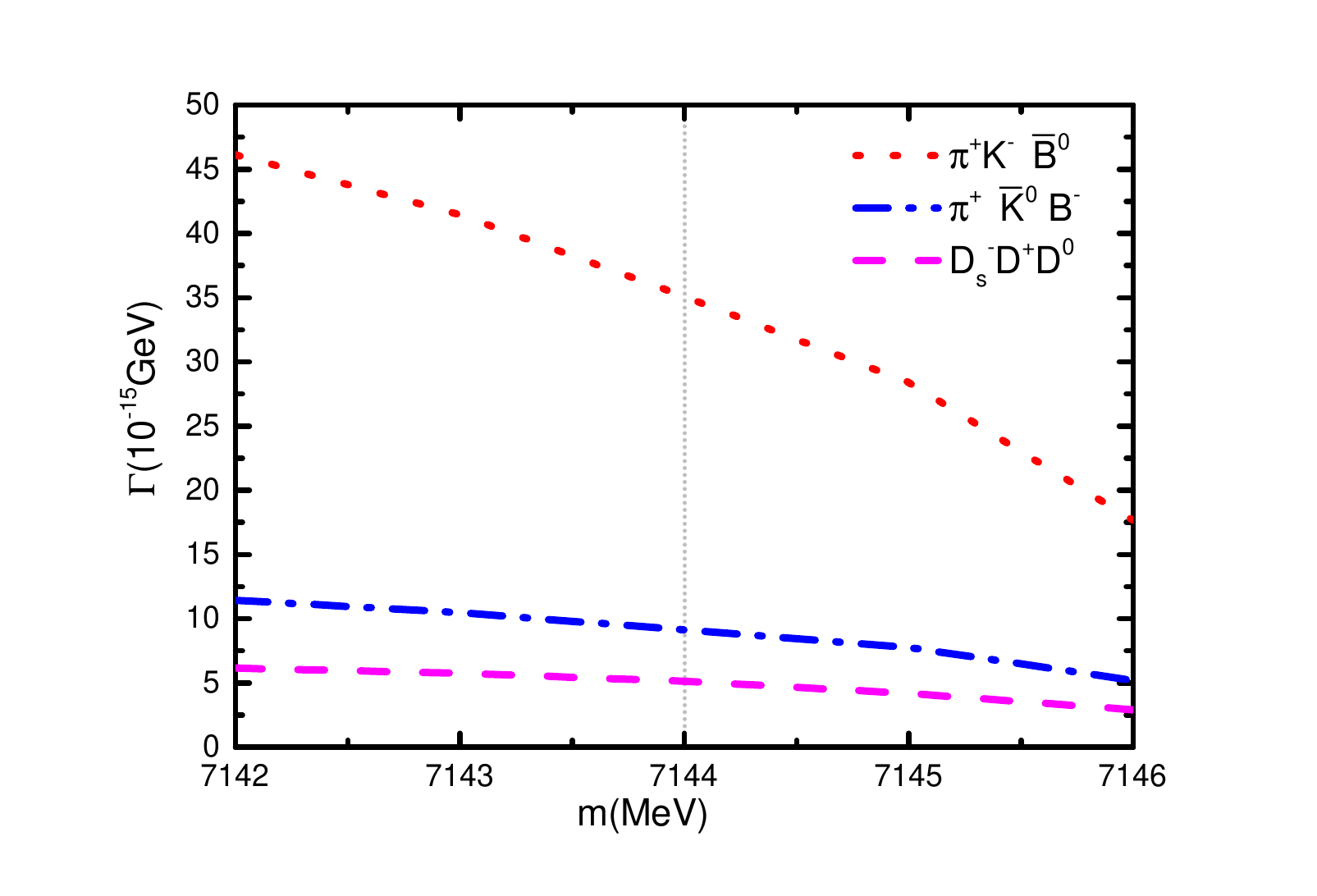}
\end{overpic}
\end{center}
\end{minipage}
} ~~~~~~~~
\subfigure[]
{
\begin{minipage}[t]{0.46\linewidth}
\begin{center}
\begin{overpic}[scale=.36]{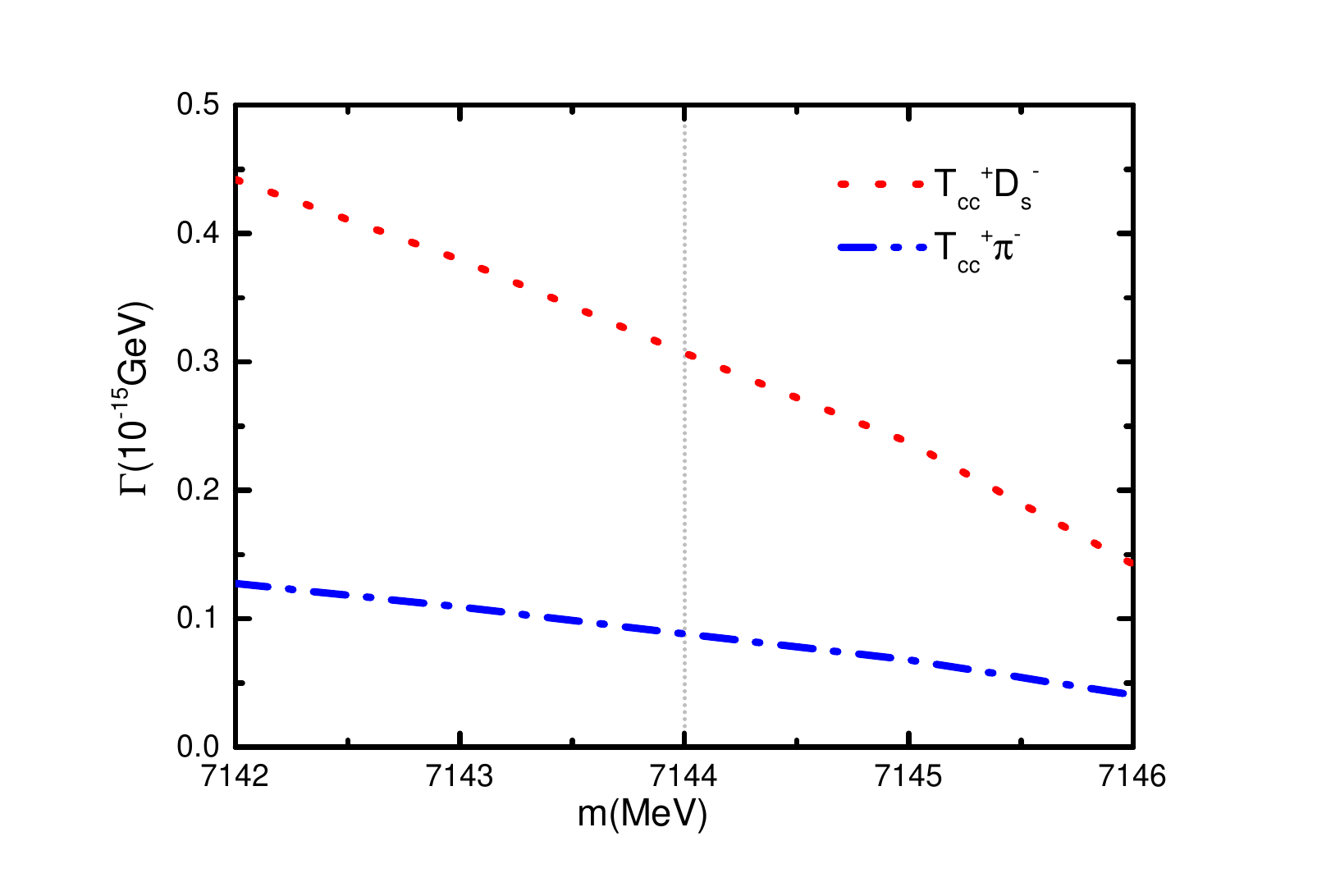}
\end{overpic}
\end{center}
\end{minipage}
}
\end{tabular}
\caption{ Partial decay widths  $T_{cb} \to D_s^- D^0 D^+$,  $T_{cb} \to \pi^+ \bar{K}^0 B^-$, and $T_{cb} \to \pi^+ K^- \bar{B}^0$ in the tree diagrams  as a function of $T_{cb}$ mass (a), and widths of partial decays $T_{cb}\to T_{cc}^+ D_{s}^- $ and $T_{cb}\to T_{cc}^+ \pi^- $ in the triangle diagrams as a function of $T_{cb}$ mass (b).       }
\label{ytox}
\end{center}
\end{figure*}

\subsection{ Decay amplitudes  }
 
With the above relevant Lagrangians, the decay amplitudes of  $T_{cb} \to D^+ D^0 D_s^-$,   $T_{cb} \to B^- \pi^+ K_s^0$,  and  $T_{cb} \to \bar{B}^0 \pi^+ K^-$   in Fig.~\ref{Fig:tree}
 can be writen as 
 \begin{eqnarray}
 i\mathcal{M}_{a,b}&=&g_{T_{cb}D\bar{B}} g_{\bar{B} \bar{D}_{s} {D}}   \frac{1}{k_0^2-m_{\bar{B}}^2}, \\  \nonumber
 i\mathcal{M}_{c,d}&=&g_{T_{cb}D\bar{B}} g_{D  K \pi}  \frac{1}{k_0^2-m_{D}^2},  
 \label{pi}
 \end{eqnarray}
 where  $k_0$ is the momentum of the $\bar{B}$ meson or $D$ meson.

Similarly, we express the decay amplitudes  of   $ T_{cb} \to T_{cc} D_s^{-} $  of Fig.~\ref{Fig:tri} 
 \begin{eqnarray}
 \label{gamma}   
 && i\mathcal{M}_{2a,2b}=g_{T_{cb}D\bar{B}} g_{T_{cc}D{D}^*} \int \frac{d^{4}q}{(2\pi)^4}  \frac{1}{k_0^2-m_{\bar{B}}^2}  \\ \nonumber  &&   
\frac{1}{q^{2}-m_{{D}}^{2}}  \frac{-g^{\mu\nu}+\frac{q_2^\mu q_2^\nu}{q_2^2}}{q_2^{2}-m_{D^*}^{2}} \mathcal{A}_{\nu}(\bar{B} \to \bar{D}_{s} {D}^*) \varepsilon_{\mu}(p_1),
 \end{eqnarray}
where $q$ and $p_1$ represent the momenta for $D$ meson and $T_{cc}$ state, and $\varepsilon_\mu$ represent the  polarization vector for the state  of spin $S=1$.

 With the weak amplitudes of $T_{cb} \to T_{cc} D_{s} $ , one can further calculate  
 the corresponding partial decay widths  as
 \begin{eqnarray}
\Gamma=\frac{1}{2J+1}\frac{1}{8\pi}\frac{|\vec{p}|}{m_{T_{cb}}^2}\bar{|\mathcal{M}|}^{2},
\end{eqnarray}
where $J$ is the total angular momentum of the initial state $T_{cb}$, the overline indicates the sum over the polarization vectors of final states, and $|\vec{p}|$ is the momentum of either final state in the rest frame of  $T_{cb}$ .  

As for the three-body decay, the partial decay widths of   $T_{cb} \to D \bar{D} \bar{D}_s$ and  $T_{cb} \to \bar{B}\pi \bar{K}$   as a function of $m_{12}^2$ and $m_{23}^2$ read
\begin{equation}
\Gamma =  \frac{1}{(2 \pi)^{3}}\frac{1}{2J+1} \int\int \frac{\overline{|\mathcal{M}|^2}}{32 m_{T_{cb}}^{3}} d m_{12}^{2} d m_{23}^{2},
\end{equation}
with $m_{12}$ the invariant mass of $\bar{D}_sD$ or $\pi\bar{K}$ and $m_{23}$ the invariant mass of $\bar{D}D$ or $\bar{B}\bar{K}$ for the $T_{cb} \to D \bar{D} \bar{D}_s$ and  $T_{cb} \to \bar{B}\pi \bar{K}$  decays, respectively.

\section{Numerical Results and Discussion}
\label{sec:Results}

\begin{table}[!h]
\caption{Masses and quantum numbers of  mesons relevant to the present work~\cite{ParticleDataGroup:2020ssz}. \label{mass}}
\begin{tabular}{ccc|ccc}
  \hline\hline
   Meson & $I (J^P)$ & M (MeV) &    Meson & $I (J^P)$ & M (MeV)   \\
  \hline  
      $\pi^{0}$ & $1(0^-)$ & $134.977$  &    $\pi^{\pm}$ & $1(0^-)$ &$139.570$  \\
    $K^{0}$ & $\frac{1}{2}(0^-)$ & $497.611$  &    $K^{\pm}$ & $\frac{1}{2}(0^-)$ & $493.677$ \\
  $D^{0}$ & $\frac{1}{2}(0^-)$ & $1864.84$  &    $D^{\pm}$ & $\frac{1}{2}(0^-)$ & $1869.66$ \\
  $D^{\ast0}$ & $\frac{1}{2}(1^-)$ & $2006.85$ &  $D^{\ast\pm}$ & $\frac{1}{2}(1^-)$ & $2010.26$
  \\
     $D_s^{\pm}$ & $0(0^{-})$ & $1968.35$ &  $T_{cc}$ & $0(1^{+})$ & $3874.74$  
     \\  $B^{\pm}$ & $\frac{1}{2}(0^-)$ & $5279.34$ &  $B^0$ & $\frac{1}{2}(0^-)$ & $5279.66$
  \\
 \hline \hline
\end{tabular}
\label{tab:masses}
\end{table}

In Table~\ref{mass}, we collect the masses and quantum numbers of the mesons relevant to the present work.   
At first, we take the contact range EFT to analyse the likely bound states with the quark content of $QQ \bar{q}\bar{q}$.  
Identifying the $T_{cc}$ as a bound state of $DD^*$, we obtain the value of $C_a+C_b=-27.26$ GeV$^{-2}$ for a cutoff $\Lambda=0.7$~GeV, and then predict a bound state below the $D^*D^*$ mass threshold  $1.58$ MeV, in agreement with the predictions of Refs.~\cite{Du:2021zzh,Albaladejo:2021vln,Dai:2021vgf,Abreu:2022sra,Montesinos:2023qbx}, which is the heavy quark spin symmetry partner of the $D^*D$ molecule. Since the two-body decay mode of the $D^*D^*$ molecule is allowed, the width of the $D^*D^*$ molecule is larger than that of the $D^*D$ molecule by two orders of magnitude~\cite{Dai:2021vgf}. As for the $D\bar{B}$ system, we turn to the light meson saturation mechanism to determine the values of $C_a$ and $C_b$. Following Refs.~\cite{Peng:2020xrf,Peng:2023lfw}, $C_{a}$ and $C_{b}$ can be written as
\begin{eqnarray}
C_{a}^{sat}&\propto& -\frac{g_{\sigma}^2}{m_{\sigma}^2}+\frac{g_{v}^2}{m_{v}^2}(1+\vec{\tau}_{1}\cdot\vec{\tau}_{2}),    \\ \nonumber
C_{b}^{sat}&\propto& \frac{f_{v}^2}{4 M^2}(1+\vec{\tau}_{1}\cdot\vec{\tau}_{2}),
\label{123}
\end{eqnarray} 
where $m_\sigma=600$ MeV, $m_v=780$~MeV,  $g_\sigma=3.4$, $g_v=2.6$, and $f_v=\kappa \cdot 2.6$ with $\kappa=2.3$ and  $M=940$~MeV~\cite{Liu:2019stu}. The product of $\tau_1 \cdot \tau_2$ is $-3$ for isospin $I=0$.   Thus the ratio of $C_b$ to $C_a$ is determined as $0.25$, and we further fix the values  of   $C_a=-21.84$ GeV$^{-2}$ and  $C_b=-5.42$ GeV$^{-2}$.  With the obtained value of $C_a$,   we  obtain a weakly  bound state below the  $D\bar{B}$ mass threshold     $2.61$ MeV, consistent with the lattice QCD simulation~\cite{Alexandrou:2023cqg}.  Finally, with the poles generated by the $D\bar{B}$ and $DD^*$ interactions,  we derive the couplings of  $g_{T_{cb}D\bar{B}}=15.16$ GeV and $g_{T_{cc}DD^*}=6.59$ GeV. In the isospin limit, we obtain the hadronic molecules couplings to the channels in particle basis,  i.e.,  $\frac{1}{\sqrt{2}}g_{T_{cb}D\bar{B}}=g_{T_{cb}D^+ B^-}=g_{T_{cb}D^0 \bar{B}^0}$ and $\frac{1}{\sqrt{2}}g_{T_{cc}DD^*}=g_{T_{cc}D^+D^{*0}}=g_{T_{cc}D^0D^{*+}}$.

In this work, we take the heavy quark limit to derive the heavy meson- heavy meson potentials, to which we assign a $15\%$ uncertainty~\cite{Nieves:2011zz,Guo:2013sya}.   
To show the impact of the heavy quark symmetry breaking on the $T_{cb}$ mass, we vary its mass from $7142$ MeV to $7146$ MeV. In Fig.~\ref{ytox}~(a), we present the partial decay widths $T_{cb} \to D_s^- D^0 D^+ $, $T_{cb} \to \pi^+ \bar{K}^0 B^-$, and $T_{cb} \to \pi^+ K^- \bar{B}^0$    as a function of the $T_{cb}$ mass.  The results show that the partial decay width of $T_{cb} \to D_s^- D^0 D^+ $ varies from $6.16 \times 10^{-15}$ GeV to $2.90 \times 10^{-15}$ GeV, and  the partial decay widths $T_{cb} \to \pi^+ \bar{K}^0 B^-$ and $T_{cb} \to \pi^+ K^- \bar{B}^0$ are in the range of $(11.43 \sim 5.22 )\times 10^{-15}$ GeV and $(4.61 \sim 1.77 )\times 10^{-14}$ GeV, respectively. In Fig.~\ref{ytox}~(b), we show the partial decay widths of $T_{cb} \to T_{cc}^+ D_{s}^-$ as a function of the $T_{cb}$ mass, which varies from $4.42 \times 10^{-16}$  to   $ 1.42 \times 10^{-16}$ GeV. Because the coupling $g_{T_{cb}D\bar{B}}$ decreases  as the $T_{cb}$ mass increases, its  partial decay widths decrease as well.

 {  In our calculation, the main uncertainty comes from the couplings of  vertices in the Feynman diagrams. 
For the couplings between molecules and their constituents, i.e., $g_{T_{cb}}$ and $g_{T_{cc}}$,    the variation of cutoff in the form factors  of scattering amplitude $T$ in Eq.~(4), varying from $0.7$ to $2$ GeV,  lead to the couplings $g_{T_{cb}}$ and  $g_{T_{cb}}$ decreasing from $15.16$ to  $13.44$ GeV and from $6.59$ to $6.17$~GeV, respectively,  resulting in  around  $10\%$ uncertainty.  The error of experimental  branching fractions of weak decays also  bring about $10\%$ uncertainties for  the couplings of weak vertices as shown in Ref.[57].    
Therefore, we estimate the 
uncertainties for the partial decay widths originating from the uncertainties of these couplings via a Monte
Carlo sampling within their $1 \sigma$ intervals.   }  
{  In  Table~\ref{partialwidths}, we show the partial decay widths and corresponding  branching fractions of $T_{cb}$ at a mass of $7144$ MeV. \footnote{ Since the widths of heavy hadrons are dominantly responsible by the weak decay of heavy flavor quarks, the widths of doubly heavy tetraquark states  are expected to be  similar to those of doubly heavy baryons.   The life time of $\Xi_{bb}$ and  $\Xi_{bc}$ are predicted to be around  $0.8$~ps and $0.28$~ps~\cite{Cheng:2019sxr}, and therefore the life time of  $T_{bc}$ is taken as $0.3$~ps or $2.2\times 10^{-9}$~MeV in this work. }  } { One can see that the error of two-body decays are larger than those of three-body decays because there exist  three couplings in  the  triangle diagrams  but two couplings in  the  tree-diagrams.      }
Our results indicate that, identifying $T_{cb}$ as a bound state of $D\bar{B}$, the decay mode  of $T_{cb} \to \pi^+ K^- \bar{B}^0$ is the largest, and therefore we suggest  to experimentally search for the $D\bar{B}$ molecule in the $ \pi^+ K^- \bar{B}^0$ mass distribution. Very recently, Ali et al., estimated the weak decay widths of $T_{cb} \to T_{cc} X$(  $X=$~$\pi^-$, $\rho^-$, and $a_1^-$ mesons)  to be the order of $10^{-15}$~GeV~\cite{Ali:2024hzp}, where the doubly heavy tetraquark states are assumed as diquark-diquark states. Such a decay mode can be produced in our mechanism. As shown in  Fig.~\ref{Fig:tri}, we replace the weak decay vertices $B^{-} \to D^{*0} D_{s}^{-} $ and $\bar{B}^{0} \to  {D}^{*+} D_{s}^{-} $  in the triangle diagrams  by  those of  $B^{-} \to D^{*0} \pi^{-} $ and $\bar{B}^{0} \to {D}^{*+} \pi^{-} $. Using the similar approach, we calculate the decay width of $T_{cb} \to T_{cc}^{+} \pi^{-}$ in the range of  $(1.24\sim 0.41)  \times 10^{-16}$~GeV as shown in Fig.~\ref{ytox}~(b), which is smaller than that of Ref.~\cite{Ali:2024hzp} by one order of magnitude. Considering the dominant decays of  $T_{cc}^{+} \to D^+ D^0 \pi^0 $ and  $T_{cc}^{+} \to  D^0 D^0 \pi^+ $, the $T_{cb}$ can also be observed in the channels of $ D^+ D^0 \pi^0 D_{s}^-$, $D^0 D^0 \pi^+D_{s}^- $, $ D^+ D^0 \pi^0 \pi^-$, and $D^0 D^0 \pi^+ \pi^- $.

\begin{table}[ttt]
\caption{ Partial decay widths and corresponding branching fractions of the $D\bar{B}$ molecule for a  mass of $7144$~MeV. \label{partialwidths}}
\begin{tabular}{cccccc}
  \hline\hline
   ~~~~ Decay mode~~~~  &~~~~ Width (GeV)~~~~  & Branching fraction $(\%)$\\
  \hline  
      $T_{cb} \to D_s^- D^0 D^+$ & $ (5.13 \pm 0.74) \times 10^{-15}$  & $0.20 \pm 0.03$ \\
            $T_{cb} \to \pi^+ \bar{K}^0 B^-$ & $ (9.15 \pm 1.51)\times 10^{-15}$  & $0.40\pm 0.07$  \\
                        $T_{cb} \to \pi^+ K^- \bar{B}^0$ & $(3.51\pm 0.58) \times 10^{-14}$  & $2.0\pm 0.3$  \\
     $T_{cb} \to T_{cc}^+ D_{s}^-$ & $(3.07 \pm 0.62) \times 10^{-16}$  & $0.0150 \pm 0.0025$    \\
          $T_{cb} \to T_{cc}^+ \pi^-$ & $(0.88 \pm 0.18 ) \times 10^{-16}$  & $0.0040\pm 0.0008$  \\
 \hline \hline
\end{tabular}
\end{table}

\section{Summary and Discussion}
\label{sum}

Since $X(3872)$ was discovered by the Belle Collaboration in 2003, many exotic states have been discovered in the experiments. Since most of them lie to the mass thresholds of a pair of conventional hadrons, the hadronic molecules are expected  to explain their internal structure. Within the molecular picture, the decay modes of  exotic states are always including the strong  decays and radiative decays. Very recently, the molecule composed by $\bar{B}$ meson and $D$ meson is only allowed to proceed via the weak decays. 

In this work, we assume  that the $D^{(*)}D^{(*)}$ system is related to the  $D^{(*)}\bar{B}^{(*)}$ system in the heavy quark symmetry. Their contact range potentials are parameterised by two parameters  $C_a$ and $C_b$. By reproducing the mass of $T_{cc}$, we determine the sum of $C_a$ and $C_b$, and then fully determine the values of  $C_a$ and $C_b$ in terms of  the ratio of  $C_a$ to $C_b$ estimated by  the light meson saturation approach. With the obtained $C_a$ and $C_b$, we predict the mass of $\bar{B}D$ molecule as $7144$~MeV, consistent with the  Lattice QCD simulations. 

Based on the decay mechanism of $DD^*$ molecule, we propose that the $D \bar{B}$ molecule decay via the weak decays of $\bar{B}$ meson or $D$ meson, i.e., $T_{cb} \to D_s^- D^0 D^+$ and  $T_{cb} \to \pi^+ \bar{K}^0 B^-$/$T_{cb} \to \pi^+ K^- \bar{B}^0$. Moreover, considering  the final state interactions, we propose the decays of  $T_{cb} \to T_{cc}^+ D_{s}^-$ and $T_{cb} \to T_{cc}^+ \pi^-$ proceeding  via the triangle diagram mechanism. Using the effective Lagrangian approach, we calculate the   partial decay widths and corresponding branching fractions  of $D\bar{B}$ molecule as shown in Table~\ref{partialwidths}.
Our calculation for the decay $T_{cb} \to T_{cc}^+ \pi^-$ is smaller than that of assuming the doubly heavy tetraquark states as the compact tetraquark states by one order of magnitude, which is an obvious signal to discriminate the nature of doubly heavy tetraquark states.      
From our calculations, we strongly suggest to experimental colleague to search for the  $D\bar{B}$ molecule in the $\pi^+ K^{-} \bar{B}^0$ mass distribution. We hope that   the present work  can simulate more studies  on the weak decays of exotic states.

\section{Acknowledgments}
M.Z.L is  grateful to  Prof. Fu-Sheng Yu  for stimulating discussions 
This work is partly supported by the National Key R\&D Program of China under Grant No. 2023YFA1606703.  M.Z.L
acknowledges support from the National Natural Science Foundation of China under Grant No.12105007.

\appendix

\bibliography{reference}

\end{document}